\documentclass[a4paper,11pt]{article}

\usepackage{rotating}
\usepackage{graphics}
\usepackage{latexsym}
\usepackage[dvips]{epsfig}
\usepackage{amssymb}
\usepackage{graphicx}
\usepackage{hyperref}
\usepackage {enumerate}
\usepackage [ruled]{algorithm}
\usepackage {algorithmic}
\usepackage {amsmath}

\usepackage{url}

\def\boxit#1{\vbox{\hrule\hbox{\vrule\kern4pt
  \vbox{\kern1pt#1\kern1pt}
\kern2pt\vrule}\hrule}}

\def\qed{\rule{1.5mm}{3mm}}

\newcommand\nc{\newcommand}

\newtheorem{theorem}{\bfseries Theorem}
\newtheorem{lemma}{Lemma}
\newtheorem{cor}{Corollary}
\newtheorem{prop}{Proposition}
\newtheorem{defn}{Definition}

\newtheorem{r-rule1}{R-Rule}

\nc{\crl}[2]{\begin{cor}\label{crl:#1} #2 \end{cor}}
\nc{\dfn}[2]{\begin{defn}\label{def:#1} #2 \end{defn}}
\nc{\lem}[2]{\begin{lemma}\label{lem:#1} #2 \end{lemma}}
\nc{\prp}[2]{\begin{prop}\label{prp:#1} #2
\end{prop}}
\nc{\thm}[2]{\begin{theorem}\label{thm:#1} #2\end{theorem}}
\nc{\fac}[2]{\begin{lemma}\label{fact:#1} #2 \end{lemma}}

\nc{\eqn}[2]{\begin{eqnarray}\label{eqn:#1} #2 \end{eqnarray}}
\nc{\fig}[4]{\begin{figure}[h]
\begin{center}
\includegraphics[width=#2\textwidth]{#4}
\end{center}
\caption{#3}\label{fig:#1}
\end{figure}}

\nc{\tbl}[3]{\begin{table}[hbt] #3 \caption{#2} \label{tab:#1}
\end{table}}

\nc{\refc}[1]{Corollary~\ref{crl:#1}}
\nc{\refd}[1]{Definition~\ref{def:#1}}
\nc{\reff}[1]{Fig.~\ref{fig:#1}}
\nc{\refl}[1]{Lemma~\ref{lem:#1}}
\nc{\refp}[1]{Proposition~\ref{prp:#1}}
\nc{\reft}[1]{Theorem~\ref{thm:#1}} \nc{\refe}[1]{(\ref{eqn:#1})}
\nc{\reftb}[1]{Table~\ref{tab:#1}}

\nc{\reffc}[1]{Fact~\ref{fact:#1}}

\nc{\pf}[1]{ \noindent \emph{Proof.} #1
 \hfill \qed\par}

\renewcommand{\title}[1]{\vspace{\fill}
\eject\addtolength{\baselineskip}{4pt}
{\bfseries\Large #1}\\[3mm]\addtolength{\baselineskip}{-4pt}}
\renewcommand{\author}[3]{\parbox[t]{75mm}
{\begin{center}{\scshape #1}\\[3mm] #2\\
 {\ttfamily #3} \end{center}}}

\long\def\invis#1{}

\begin{document}

\begin{center}

\title{An Improved Upper Bound for SAT
%\footnote{~{\bf Technical report 2013-00?, October ??, 2013.}  A preliminary version of this paper was presented at the 24th international symposium on algorithms and computation (ISAAC 2013).}
}

\author{Huairui Chu}
 {School of Computer Science and Engineering,
University of Electronic Science and Technology of China, China,
}{a1444933023@163.com}
\author{Mingyu Xiao}{School of Computer Science and Engineering,
University of Electronic Science and Technology of China, China,
}{myxiao@gmail.com}
\author{Zhe Zhang}{School of Computer Science and Engineering,
University of Electronic Science and Technology of China, China,
}{2017060106011@std.uestc.edu.cn}
% \footnotetext[1]{Technical report 2013-006, November ??, 2013 }
%
%http://www-or.amp.i.kyoto-u.ac.jp/members/nag/Technical_report/TR2012-002.pdf
\end{center}

\begin{abstract}
We show that the CNF satisfiability problem can be solved $O^*(1.2226^m)$ time, where $m$ is the number of clauses in the formula, improving the known upper bounds $O^*(1.234^m)$ given by Yamamoto 15 years ago and $O^*(1.239^m)$ given by Hirsch 22 years ago.
By using an amortized technique and careful case analysis, we successfully avoid the bottlenecks in previous algorithms and get the improvement.

% \noindent {\bf Key words.}  \ \
%Exact Algorithm, Independent Set, Graph, Polynomial-space, Branch-and-reduce, Measure-and-conquer,
%Divide-and-conquer
\end{abstract}

\section{Introduction}

The problem of testing the satisfiability of a propositional formula in conjunctive normal form (CNF), denoted by SAT, is one of the most fundamental problems in computer science. It is the first problem proved to be NP-complete \cite{COOK1971} and plays an important role in computational complexity and artificial intelligence~\cite{AGuide1979}. To make the problem tractable, a large number of references studied it from the view of heuristic algorithms, approximation algorithms, randomized algorithms, and exact algorithms.
In this paper, we study exact algorithms for SAT with guaranteed theoretical running time bounds.

\subsection{Related Works}
To evaluate the running time bound, there are three frequently used measures: the number of
variables $n$, the number of clauses $m$, and the length of the whole input $L$.
The trivial algorithm to check all possible assignments runs in $O^*(2^n)$ time\footnote{The notation $O^*$ suppresses all polynomially bounded factors. For two functions $f$ and $g$, we write $f(n)=O^*(g(n))$ if $f(n)=g(n)n^{O(1)}$.}.
A nontrivial bound better than $O^*(2^n)$ was obtained in \cite{FirstGeneral}, which is $O^*(2^{n(1-2\sqrt{1/n\log m})})$. Later better upper bounds were introduced in \cite{BetterGeneral} and \cite{Schuler}.
However,  no algorithm with running time bound $O^*(c^n)$ for some constant $c<2$ was found, despite decades of hard work. The nonexistence of these algorithms is known as the Strong Exponential Time Hypothesis (SETH) \cite{SETH}.
On the other hand, for a restricted version, the $k$-SAT problem (where each clause in the CNF-formula contains at most $k$ literals), a series of significant results have been developed. A branch-and-bound technique was introduced in \cite{MonienFirst} and \cite{DantsinFirst}, which can solve $k$-SAT in $O^*((\alpha_k)^n)$ time where $\alpha_k$ is the largest root of the function $x=2-1/x^{k-1}$. After this, a series of improvements on the upper bounds for $k$-SAT have been made. Most of them are based on derandomization, such as the $O^*(2^{(1-1/2k)n})$ bound in \cite{PPZ} and the $O^*((2-2/(k+1))^n)$ bound in \cite{Schoning}. Recently a new randomized algorithm for $k$-SAT with better running time bound was introduced \cite{19kSAT}.

When the length of the input $L$ is taken as the measure, from the first algorithm with running time bound $O^*(1.0927^L)$ by Gelder~\cite{Van}, the result was improved frequently. Let us quote the bound $O^*(1.0801^L)$ by Kullmann~\cite{Kullmann},
$O^*(1.0758^L)$ by Hirsch~\cite{Hirsch}, $O^*(1.074^L)$ by Hirsch~\cite{Hirsch2}, and $O^*(1.0663^L)$ by Wahlstr{\"{o}}m~\cite{Wahlstrom}. Currently, the best known bound was $O^*(1.0652^L)$ obtained by Chen and Liu~\cite{Chen}.

Another important measure is the number of clauses $m$. Monien and Speckenmeyer~\cite{Monien1980} gave an $O^*(1.260^m)$-time
algorithm in 1980, which was improved to $O^*(1.239^m)$ by Hirsch \cite{Hirsch} in 1998.
Then it  took seven years for Yamamoto to slightly improve Hirsch's bound to $O^*(1.234^m)$ \cite{Yamamoto}.
% and then to $O^*(1.234^m)$ by Yamamoto \cite{Yamamoto}.
%It took seven years for Yamamoto to slightly improve Hirsch's bound.
In this paper, we will significantly improve Yamamoto's bound obtained 15 years ago.  Previous and our results are listed in  Table~\ref{tab:plain}.

\begin{table}[h!]
	\caption{Previous and our upper bounds for SAT}
	\centering
	
	\begin{tabular}{lc}
		\hline
		running times  &  references \\
		\hline
		$O^*(1.260^m)$       & \cite{Monien1980}   \\
		$O^*(1.239^m)$       & \cite{Hirsch}                    \\
		$O^*(1.234^m)$       & \cite{Yamamoto}                  \\
		$O^*(1.2226^m)$       & \textbf{This paper}              \\
		\hline
	\end{tabular}
	
	\label{tab:plain}
\end{table}

\subsection{The Techniques}
%Most previous SAT algorithms can be classified into two classes: one is randomized algorithm and the other one is branch-and-search algorithm.
%For SAT measured by the number of variables $n$, most fast algorithms, such as the algorithms in \cite{BetterGeneral,PPZ,19kSAT}, use randomization techniques.
%For SAT measured by the number of clauses $m$,
All algorithms in Table~\ref{tab:plain} are branch-and-search algorithms.
The branch-and-search idea is simple and practical: we iteratively branch on a literal into two branches by letting it be 1 or 0.
Consider an $(a,b)$-literal (a literal such that itself appears in $a$ clauses and the negation of it appears in $b$ clauses). In the branching where the literal is assigned 1,
we can reduce $a$ clauses; in the branching where the literal is assigned 0, we can reduce $b$ clauses.
We hope that the values of $a$ and $b$ are larger, so that we can reduce the instance to a greater extent. There are several developed techniques to deal with $(a,b)$-literals
with small values of $a$ and $b$, say one of them is at most 2. Thus the worst case will become to branch on a $(3,3)$-literal, in which we can only get a  branching vector of $(3,3)$
and a branching factor 1.2600. We get the bound of  $O^*(1.260^m)$~\cite{Monien1980}. It seems that branching on $(3,3)$-literals is unavoidable.
Hirsch~\cite{Hirsch} showed that after branching on a $(3,3)$-literal we can always branch with a branching vector at least $(4,3)$ or $(3,4)$ subsequently. Combing the bad branching vector $(3,3)$ with the good branching vector $(4,3)$ or $(3,4)$, he got a better worst-case and then improved the running time bound to $O^*(1.239^m)$.
Yamamoto \cite{Yamamoto} further showed that the worst cases in Hirsch's algorithm would not always happen: we can further branch with $(4,3)$ or $(3,4)$ at the third level, i.e., after branching with $(4,3)$ or $(3,4)$ after branching with $(3,3)$. Yamamoto considered more levels of the branching but could only slightly improve the bound to $O^*(1.234^m)$. The improvement is very slow, and we seem to have reached the bottleneck.

Our algorithm is still a branch-and-search algorithm, following the main framework in the previous algorithms. We still can not avoid branching on $(3,3)$-literals,
otherwise the worst case would be to branch on $(3,4)$-literals or $(4,3)$-literals and the bound would be improved to $O^*(1.2208^m)$.
We also show that after branching on a $(3,3)$-literal we can further branch with better branching vectors.
However, the traditional analysis to combine several levels of branchings into a big branching is somewhat complicated and limited.
To exhibit the relations among good and bad branchings in our algorithm and also to use as many good branchings as possible to even out the bad ones,
we will use an amortized technique to analyze the running time bound.
To get the claimed result, we also need to use some new reduction and branching rules and deep analysis of the structure.

\section{Preliminaries}

Let $V=\{x_1,x_2,\dots,x_n\}$ denote a set of $n$ boolean \emph{variables}. For each variable $x_i$ $(i=1,2,3,\dots, n)$,
a \emph{literal} is either $x_i$ or the negation of it $\overline{x_i}$ (we use $\overline{x}$ to denote the negation of a literal $x$, and then $\overline{\overline{x}}=x$). A \emph{clause} on $V$ is a set of literals on $V$ without a negation of any literal in it, which means $x$ and $\overline{x}$ cannot be contained simultaneously in a clause for any variable $x\in V$. A \emph{CNF-formula} on $V$ is a sequence of clauses $\mathcal{F}=\{C_1,C_2,C_3,\dots,C_m\}$. We will use $m_{\mathcal{F}}$ to denote the number of clauses in $\mathcal{F}$.
An \emph{assignment} for $V$ is a map $A: V\rightarrow\{0,1\}$.
%Any variable $x_i$ is \emph{satisfied} if and only if $A(x_i)=1$ and $\overline{x_i}$ is \emph{satisfied} if and only if $A(x_i)=0$.
A clause $C_j$ on $V$ is \emph{satisfied} by $A$ if and only if there exists a literal $x$ in $C_j$ such that $A(x)=1$.
%which means an empty clause cannot be satisfied.
A CNF-formula is satisfied by an assignment $A$ if and only if each clause in it is satisfied by $A$. An assignment $A$ that makes a CNF-formula $\mathcal{F}$ satisfied is called a \emph{satisfying assignment} for $\mathcal{F}$.
Given a CNF-formula $\mathcal{F}$ on a set of variables $V$, the SAT problem is to check the existence of a satisfying assignment
for $\mathcal{F}$.
%In the following of the paper, if the variable set $V$ or the formula $\mathcal{F}$ is clear from the context, we may simply omit the description.

The \emph{degree} of a literal $x$ in $\mathcal{F}$ is the number of clauses in $\mathcal{F}$ containing it.
The \emph{total degree} of a literal $x$ is the degree of $x$ plus the degree of $\overline{x}$.
If the degree of $x$ is $a$ (resp., at least $a$ or at most $a$) and the degree of $\overline{x}$ is $b$, we say $x$ is an \emph{$(a,b)$-literal} (resp., an \emph{$(a^+,b)$-literal} or an \emph{$(a^-,b)$-literal}). Similarly, we can define $(a,b^+)$-literal, $(a,b^-)$-literal, $(a^+,b^+)$-literal, $(a^-,b^-)$-literal and so on.
Note that a literal $x$ is an $(a,b)$-literal if and only if $\overline{x}$ is a $(b,a)$-literal.
A clause containing exactly $c$ literals is called a \emph{$c$-clause}.
%We say that a formula $\mathcal{F}$ contains a literal $x$ or there is a literal $x$ in $\mathcal{F}$ if and only if the degree of $x$ is positive.
%Obviously, if $\mathcal{F}$ contains an $(a,b)-literal(b\geq 1)$, $\mathcal{F}$ also contains a $(b,a)-literal$.
%Degree, (a,b)-literal, c-clause, containing a literal
A pair of literals $x$ and $y$ is called a \emph{coincident pair} if there are at least two clauses containing them simultaneously.

%Coincidence is a concept defined by Yamamoto. If there are two literals $x$, $y$ and there are two clauses containing them simultaneously, we say they are a coincident pair.

Our algorithm will first apply reduction rules to reduce the instance and then apply branching rules to search for a solution when the instance can not be further reduced. Next, we first introduce the reduction rules.

\section{Reduction Rules}
We have five reduction rules. The first two are easy to observe and used in the literature \cite{Davis}.

\begin{r-rule1}\textbf{\emph{(Elimination of 1-clauses and pure literals)}}  If the CNF-formula contains a $1$-clause $\{x\}$ or an $(a,0)$-literal $x$ with $a>0$, assign $x=1$.
\end{r-rule1}
%We will use $R_1(\mathcal{F},x)$ to denote the application of R-Rule 1 on a $1$-clause $\{x\}$ in CNF-formula $\mathcal{F}$.

\begin{r-rule1} \textbf{\emph{(Elimination of subsumptions)}} If the CNF-formula contains two clauses $C$ and $C'$ such that $C\subseteq C'$, then delete $C'$.
\end{r-rule1}
%We will use $R_2(\mathcal{F},x)$ to denote the application of R-Rule 1 on an $(a,0)$-literal $x$ in CNF-formula $\mathcal{F}$.

The following proposition is known as the resolution technique in the literature, which was first proved in~\cite{JARobinson}, and then used in many SAT algorithms.

\dfn{resolution}{ \textbf{\emph{(Resolution on a variable)}}
	Let $\mathcal{F}$ be a CNF-formula containing a variable $x$.
	Let $E_1,E_2,\dots, E_a$ be the clauses containing $x$ and $D_1,D_2,\dots, D_b$ be the clauses containing $\bar{x}$.
	\emph{Resolving} on variable $x$ is to construct a new CNF-formula $\mathcal{F}_{\setminus x}$ by the following method:
	for each $i\in \{1,2,\dots, a\}$ and $j\in \{1,2,\dots,b\}$, add the clause
	$F_{ij}=E_i\cup D_j\setminus \{x,\bar{x}\}$ to the formula if it does not contain both a literal and the negation of it;
	delete $E_i$ ($i\in \{1,2,\dots, a\}$) and $D_j$ ($j\in \{1,2,\dots,b\}$) from the formula.
}

We may always use $\mathcal{F}_{\setminus x}$ to denote the CNF-formula after resolving a variable $x$ in $\mathcal{F}$.

\prp{resolution-eq}{ \emph{\cite{JARobinson}}
	Let $\mathcal{F}$ be a CNF-formula containing a variable $x$ and $\mathcal{F}_{\setminus x}$ be the CNF-formula after resolving on variable $x$.
	Then $\mathcal{F}$ has a satisfying assignment if and only if $\mathcal{F}_{\setminus x}$ does.
}

\begin{r-rule1} \textbf{\emph{(Resolving on some variables)}} If there is an $(a,b)$-literal $x$ such that $a=1$ and $b\geq 1$ or $a=2$ and $b=2$,
	then resolve $x$ in $\mathcal{F}$, i.e., replace $\mathcal{F}$ with $\mathcal{F}_{\setminus x}$.
\end{r-rule1}

We also introduce a simple but powerful concept, based on which we can design several reduction rules. %This rule is discovered by Hirsch.

\dfn{Autarkic-sets}{ \textbf{\emph{(Autarkic sets)}}
	A set $X$ of literals is called an \emph{autarkic set} if each clause containing a negation of a literal in $X$ also contains a literal in $X$.
}

\lem{Autarkiceq}{
	If a CNF-formula $\mathcal{F}$ has a satisfying assignment, then it has a satisfying assignment where all literals in an autarkic set are assigned 1.
}

\pf{
	If we assign 1 to all literals in an autarkic set $X$, then any clause containing either a literal in $X$ or a negation of a literal in $X$ is satisfied, since each clause containing a negation of a literal in $X$ also contains a literal in $X$. Any other assignment of literals in $X$ can only satisfy a subset of these clauses. So we can simply assign 1 to all literals in $X$.
}\medskip

The following reduction rule was firstly used in \cite{Hirsch}. It is an application of a special autarkic set.
\begin{r-rule1} \emph{\cite{Hirsch}} If each clause containing a $(2,3^+)$-literal also contains a $(3^+,2)$-literal, assign $1$ to each $(3^+,2)$-literal.
\end{r-rule1}

Our algorithm also needs to eliminate another kind of autarkic sets.

\begin{r-rule1} Let
	%$C=\{x| x \text{ is a } (4,3)\text{-literal and there is a clause containing } x \text{ and a } (3,3^+)\text{-literal}\}$.
	$X$ be the set of $(4,3)$-literals $x$ such that there is a clause containing both $x$ and a $(3,3^+)$-literal.
	If each clause containing a negation of a literal in $X$ also contains a $(4,3)$-literal, assign $1$ to each literal in $X$.
\end{r-rule1}

Each clause containing a negation of a literal $x\in X$ also contains a $(4,3)$-literal $y$. Since $\bar{x}$ is a $(3,4)$-literal, we know that $y$ is also in $X$. Thus $X$ is an autarkic set. In this reduction rule, the requirement of `a clause containing both $x$ and a $(3,3^+)$-literal' plays no role in establishing $X$ to be an autarkic set.
This requirement is used to identify a particular subset of  $(4,3)$-literals, which will be useful in our analysis.

%In R-Rule 1 and 2 we assign $1$ to a special literal. We say applying R-Rule 1 or 2 on $x$ if the assigned literal is $x$.
%In R-Rule 3, if a literal $x$ is to be resolved, we say applying R-Rule 3 on $x$.

%Based on the above analysis, the following property is obvious.
\lem{lemma-rr1}{
	After applying any of the above reduction rules, the satisfiability of the formula does not change.
	Except for the application of R-Rule 3 on a $(2,2)$-literal
	where the number of clauses does not increase, each application of other reduction rules decreases the clause number by at least 1.
}

\dfn{reduced-formulas}{ \textbf{\emph{(Reduced formulas)}}
	A formula is called \emph{reduced} if none of the five reduction rules can be applied on the formula.
}

For an instance $\mathcal{F}$, we will use $R(\mathcal{F})$ to denote the resulting reduced formula after iteratively applying the reduction rules on $\mathcal{F}$.

\lem{reduce-polytime}{
	Given a formula, we can apply the five reduction rules in polynomial time to change it to a reduced formula.
}
\pf{
	It is easy to see that each reduction rule can be applied in polynomial time. Since each reduction rule either assigns a literal to 1 or resolve a variable, we know that we can apply
	at most $n$ times of reduction rules. Thus, the total running time is bounded by a polynomial.
}\medskip

\lem{lem-d4}{
	Let $\mathcal{F}$ be a reduced formula. Then there is no 1-clause, $(2,2)$-literal or $(1^-,a)$-literal with $a\geq 1$ in $\mathcal{F}$.
	Furthermore, the total degree of any literal in $\mathcal{F}$ is at least $5$.
}
\pf{
	If there is a $(0,a)$-literal, then R-Rule 1 would be applicable. If there is a $(1,a)$-literal, then R-Rule 3 would be applicable. If there is a $(2,2)$-literal,
	then R-Rule 3 would be applicable. If there is a 1-clause, then R-Rule 1 would be applicable. All these contradict the fact that $\mathcal{F}$ is reduced.
	
	If a literal has a total degree at most 4, then it must be a $(2,2)$ or $(1^-,a)$ or $(a,1^-)$-literal. For the last case, the negation of the literal is a $(1^-,a)$-literal.
}\medskip

\section{Branch-and-Search Paradigms}
Our algorithm will first apply our reduction rules to reduce the instance. When no reduction rule can be applied anymore, we will branch to search for a solution. Our branching rule is simple. We take a literal $x$ and branch on it into two sub-instances. In one sub-instance we assign $x=1$ and in the other one we assign $x=0$, i.e, we get two sub instances
$\mathcal{F}_x$ and $\mathcal{F}_{\bar{x}}$. Selecting different literals to branch will lead to different algorithms. We want to select `good' literals to branch on such that the size of the sub instances can be reduced fast.

We use the number $m$ of clauses to evaluate the size of the formula.
Assume the number of clauses of the current instance is $m$. If a branching operation branches into $l$ sub-branches such that the number of clauses in the $i$-th sub-instance decreases by at least $c_i$, we say this operation branches with a \emph{branching vector} $(c_1, c_2, \dots, c_l)$. The largest root of the function $f(x) = 1 - \sum_{i=1}^{l} x^{-c_i}$ is called the \emph{branching factor}. If $\gamma$ is the maximum branching factor among all branching factors in an algorithm, then the running time of the algorithm is bounded by $O^*(\gamma^m)$. More details about the analysis and how to solve recurrences can be found in the monograph \cite{exactalgorithm}.
The following property is frequently used in the paper: for two branching vectors $C=(c_1, c_2, \dots, c_l)$ and $B=(b_1, b_2, \dots, b_l)$, if it holds that $c_i\geq b_i$ for each $i$, then we say $B$ \emph{covers} $C$.
The corresponding branching factor of a branching vector $C$ is not greater than the corresponding branching factor of a branching vector that covers $C$.

\subsection{Good formulas \& bad formulas}
Similar to the technique used by  Niedermeier and Rossmanith to solve the 3-hitting set problem~\cite{Niedermeier}, we also classify formulas in our algorithm into two classes: good formulas and bad formulas.
For good formulas, we may be able to branch with good branching vectors. For bad formulas, we may only be able to get bad branching vectors. 
We will show that bad formulas will not appear frequently. Then we can use an amortized analysis to get better branching vectors. 
%This method is similar to one used to solve 3-HITTING SET problem by Niedermeier and Peter Rossmanith \cite{Niedermeier}.
% and the similar thought was generalized by Wahlstr$\"{o}$m \cite{wahlstrom2}.
To make the amortized analysis easy to follow, we will use the substitution method to prove our bounds.
The precise definitions of good and bad formulas are given below.

\dfn{good-bad-formulas}{ \textbf{\emph{(Good formulas \& bad formulas)}}
	A formula $\mathcal{F}$ is a \emph{bad} formula if and only if the following four conditions are satisfied
	\begin{enumerate}[(1)]%TODO may use (a)
		\item $\mathcal{F}$ only contains $(3,3)$-literals, $(3,4)$-literals and $(4,3)$-literals.
		\item There is no coincident pair.
		\item There is no $2$-clause.
		\item There is no clause containing a $(4,3)$-literal and a $(3,3^+)$-literal simultaneously.
	\end{enumerate}
	A formula is \emph{good} if it is not a bad formula.
}

\subsection{The algorithm and its analysis}

The main steps of our algorithm are listed in Algorithm~\ref{alg:algorithm}. The precise descriptions and analysis of lines 11 and 14 are delayed to Sections~\ref{sec_bad} and~\ref{sec_good}.

\begin{algorithm}[h!]
	\caption{SAT($\mathcal{F}$)}
	\label{alg:algorithm}
	%\textbf{Input}: A reduced CNF-Formula $\mathcal{F}$\\
	%\textbf{Output}: True or False
	\begin{algorithmic}[1]
		\IF{\{$\mathcal{F}$ is not reduced\}}
		\STATE Iteratively apply our reduction rules to reduce it.
		\ENDIF
		\IF{\{$\mathcal{F}$ is empty\}}
		\STATE Return true.
		\ENDIF
		\IF{\{$\mathcal{F}$ contains an empty clause\}}
		\STATE Return false.
		\ENDIF
		\IF{\{$\mathcal{F}$ is a bad formula\}}
		\STATE Apply branching rules in Sec.~\ref{sec_bad} to search for a solution.
		\ENDIF
		\IF{\{$\mathcal{F}$ is a good formula\}}
		\STATE Apply branching rules in Sec.~\ref{sec_good} to search for a solution.
		\ENDIF
	\end{algorithmic}
\end{algorithm}

Recall that, for an instance $\mathcal{F}$, $R(\mathcal{F})$ is the resulting reduced instance after applying the reduction rules on $\mathcal{F}$, and $m_{\mathcal{F}}$ is the number of clauses in $\mathcal{F}$.
We have the following important lemmas, which are the base for us to establish the running time bound.

\lem{consistency}{
	Let $\mathcal{F}$ be a CNF-formula. It holds that $m_{R(\mathcal{F})}\leq m_{\mathcal{F}}$. Furthermore,
	if $\mathcal{F}$ is good, then either $R(\mathcal{F})$ is good or $m_{R(\mathcal{F})}\leq m_{\mathcal{F}}-1$.
}

\pf{
		By \refl{lemma-rr1}, we have that $m_{R(\mathcal{F})}\leq m_{\mathcal{F}}$.
		Next, we assume that $\mathcal{F}$ is good.
		
		If $R(\mathcal{F})=\mathcal{F}$, obviously $R(\mathcal{F})$ is good. So we assume that some R-Rules are applied. By \refl{lemma-rr1}, we know that if $m_{R(\mathcal{F})}=m_{\mathcal{F}}$ then only R-Rule 3 is applied on $(2,2)$-literals.
		For any $\mathcal{F}'$ with a $(2,2)$-literal $x$ in it, we show that after applying R-Rule 3 on $x$ the resulting instance $\mathcal{F}'_{\setminus x}$ is good. Let the two clauses containing $x$ in $\mathcal{F}'$ be $D_1$ and $D_2$, the two clauses containing $\bar{x}$ be $E_1$ and $E_2$. If $m_{\mathcal{F}'}=m_{\mathcal{F}'_{\setminus x}}$, then all $E_{ij}=D_i\cup E_j \setminus \{x,\bar{x}\}$ for each $1\leq i,j\leq 2$ are in $\mathcal{F}'_{\setminus x}$. If one of $D_1$, $D_2$, $E_1$ and $E_2$ contains at least three literals, then we will get some coincident pair. Otherwise, each $E_{ij}$ is a $2$-clause. For any case, $\mathcal{F}'_{\setminus x}$ is good.
}

\lem{branch-bad}{
	If the formula $\mathcal{F}$ is reduced and bad, then our algorithm can branch with
	either a branching vector covered by $(3,4)$ or $(4,3)$,
	or a branching vector $(3,3)$ such that the formula in each branch is good.
}

\lem{branch-good}{
	If the formula to branch is reduced and good, then our algorithm can branch with
	either a branching vector covered by one of
	$(3,5)$, $(5,3)$, and $(4,4)$,
	or a branching vector $(3,4)$ or $(4,3)$ such that the formula in each branch is good.
}

The proof of \refl{branch-bad} and \refl{branch-good} are given in Sections~\ref{sec_bad} and~\ref{sec_good}, respectively. Next, we prove the running time bound of the algorithm based on
\refl{consistency}, \refl{branch-bad}, and \refl{branch-good}.

\thm{thm-final}{
	SAT can be solved in $O^*(1.2226^m)$ time.
}

\pf{
	We use $T(\mathcal{F})$ to denote the size of the search tree generated by the algorithm running on an instance $\mathcal{F}$. We only need to prove that $T(\mathcal{F})=O(1.2226^{m_{\mathcal{F}}})$.
	%We use  $T_g(m)$ (resp., $T(T_b(m))$) to denote the worst size of the search tree running on any good formula (resp., bad formula) with at most $m$ clauses.
	To prove the theorem, we will show that there are two constants $c_1=2$ and $c_2=c_1/0.9136$ such that
	\begin{equation} T(\mathcal{F})\leq c_1 1.2226^{m_{\mathcal{F}}}-1,~~\mbox{if $\mathcal{F}$ is good,} \label{the1}\end{equation}
	and
	\begin{equation} T(\mathcal{F})\leq c_2 1.2226^{m_{\mathcal{F}}}-1,~~\mbox{if $\mathcal{F}$ is bad.}\label{the2}\end{equation}
	
	First of all, we show that we can assume $\mathcal{F}$ is a reduced instance without loss of generality.
	If the current instance $\mathcal{F}$ with $m$ clauses is not a reduced one, our algorithm will apply reduction rules on it
	to get a reduced instance $\mathcal{F}^*$ with $m^*$ clauses. To prove that (\ref{the1}) and (\ref{the2}) hold for $\mathcal{F}$, we only need to prove that (\ref{the1}) and (\ref{the2}) hold for $\mathcal{F}^*$. The reason is based on the following observations. If both of $\mathcal{F}$ and $\mathcal{F}^*$ are bad or good, then it holds that $c_i 1.2226^{m_{\mathcal{F}^*}} \leq c_i 1.2226^{m_{\mathcal{F}}}$ since $m_{\mathcal{F}^*}\leq m_{\mathcal{F}}$ by \refl{consistency}.
	If $\mathcal{F}$ is bad and $\mathcal{F}^*$ is good, then it holds that $c_1 1.2226^{m_{\mathcal{F}^*}} \leq c_2 1.2226^{m_{\mathcal{F}}}$.
	If $\mathcal{F}$ is good and $\mathcal{F}^*$ is bad, then it still holds that $c_2 1.2226^{m_{\mathcal{F}^*}} \leq c_1 1.2226^{m_{\mathcal{F}}}$ because now we have  $m_{\mathcal{F}^*}\leq m_{\mathcal{F}}-1$ by \refl{consistency} and then $c_1< 1.2226 c_2$.
	
	Next, we simply assume that the instance $\mathcal{F}$ is reduced and use $\mathcal{F}_1$ and $\mathcal{F}_2$ to denote the two sub instances generated by our branching operations.
	We use the substitution method to prove (\ref{the1}) and (\ref{the2}).
	
	Assume that $T(\mathcal{F})\leq c_i 1.2226^{m_{\mathcal{F}}}-1$ (where $c_i=c_1$ if $\mathcal{F}$ is good and $c_i=c_2$ if $\mathcal{F}$ is bad) holds for all instances $\mathcal{F}$ with less than $m$ clauses. We show that it also holds for instances with $m$ clauses.

	First, we consider the case where $\mathcal{F}$ is bad. According to  \refl{branch-bad}, there are two cases. For the first case of branching with a vector $(3,4)$ or $(4,3)$, we have that
	\begin{align}
	T(\mathcal{F})&=T(R(\mathcal{F}_1))+T(R(\mathcal{F}_2))+1\nonumber
	\\ &\leq c_2 1.2226^{m_{R(\mathcal{F}_1)}}+ c_2 1.2226^{m_{R(\mathcal{F}_2)}}-1 \nonumber
	\\ & ~~~~~~~~~~~~~~~~~~~~~~~~~~~~\text{(by the assumption and $c_1< c_2$)} \nonumber
	\\ &\leq c_2 1.2226^{m_{\mathcal{F}}-3}+c_2 1.2226^{m_{\mathcal{F}}-4}-1\nonumber
	\\ &\leq c_2 1.2226^{m_{\mathcal{F}}}-1.\nonumber
	\end{align}
	For the second case of branching with a vector $(3,3)$, the two sub instances are good, we have that
	\begin{align}
	T(\mathcal{F})&=T(R(\mathcal{F}_1))+T(R(\mathcal{F}_2))+1\nonumber
	\\ &\leq c_1 1.2226^{m_{\mathcal{F}}-3}+c_1 1.2226^{m_{\mathcal{F}}-3}-1\nonumber
	\\ &\leq c_2 1.2226^{m_{\mathcal{F}}}-1.\nonumber
	\end{align}%TODO The number?
	
	Second, we consider the case where $\mathcal{F}$ is good. According to \refl{branch-good}, there are two cases.
	
	In the first case, the branching vector is $(3,5)$ or $(5,3)$ or $(4,4)$.
	%analogously we can prove that $T(\mathcal{F})\leq c_1 1.2226^{m_{\mathcal{F}}}-1$. We omit the details here during the limited space.
	If it is $(3,5)$ or $(5,3)$, we have that
	\begin{align}
	T(\mathcal{F})&=T(R(\mathcal{F}_1))+T(R(\mathcal{F}_2))+1\nonumber
	\\ &\leq c_{i_1} 1.2226^{m_{\mathcal{F}}-3}+c_{i_2} 1.2226^{m_{\mathcal{F}}-5}-1\nonumber
	\\ &\leq c_{2} 1.2226^{m_{\mathcal{F}}-3}+c_{2} 1.2226^{m_{\mathcal{F}}-5}-1\nonumber
	\\ &\leq c_1 1.2226^{m_{\mathcal{F}}}-1,\nonumber
	\end{align}
	where $c_{i_1}, c_{i_2}\in \{1,2\}$.
	If the branching vector is $(4,4)$, we have that
	\begin{align}
	T(\mathcal{F})&=T(R(\mathcal{F}_1))+T(R(\mathcal{F}_2))+1\nonumber
	\\ &\leq c_{i_1} 1.2226^{m_{\mathcal{F}}-4}+c_{i_2} 1.2226^{m_{\mathcal{F}}-4}-1\nonumber
	\\ &\leq 2c_{2} 1.2226^{m_{\mathcal{F}}-4}-1\nonumber
	\\ &\leq c_1 1.2226^{m_{\mathcal{F}}}-1,\nonumber
	\end{align}
	where $c_{i_1}, c_{i_2}\in \{1,2\}$.

	For the second case of branching with a vector $(3,4)$ or $(4,3)$ such that the two sub instances are good, we have that
	\begin{align}
	T(\mathcal{F})&=T(R(\mathcal{F}_1))+T(R(\mathcal{F}_2))+1\nonumber
	\\ &\leq c_1 1.2226^{m_{\mathcal{F}}-3}+c_1 1.2226^{m_{\mathcal{F}}-4}-1\nonumber
	\\ &\leq c_1 1.2226^{m_{\mathcal{F}}}-1.\nonumber
	\end{align}
	
	We have proved that %$T(\mathcal{F})\leq c_i 1.2226^{m_{\mathcal{F}}}-1$ (where $c_i=c_1$ if $\mathcal{F}$ is good and $c_i=c_2$ if $\mathcal{F}$ is bad) for any instance $\mathcal{F}$.
	(\ref{the1}) and (\ref{the2}) hold for $\mathcal{F}$.
	Thus, it holds that $T(\mathcal{F})=O(1.2226^{m_{\mathcal{F}}})$, no matter $\mathcal{F}$ is good or bad.
}

\section{Some Properties}
Before giving the detailed steps of the branching operations,
%In the next section we are going to complete the detailed proof for Lemma~\ref{branch_bad} and \ref{branch_good}.
we give some properties that will be used to simplify our presentation and analysis.

In a branching operation, we need to analyze the branching vector, i.e., the number of clauses decreased in each branching.
Sometimes we can get a branching vector good enough for our analysis, such as branching vectors $(4,4)$, $(3,5)$ and $(5,3)$.
Sometimes the branching vector is not good enough and we still need to prove the remaining formulas are good, which will allow
us to use amortization. Usually, we will fall in one of the following two cases:
\begin{enumerate}
	\item Some variables are assigned values (including applying R-Rule 1) and then some clauses are deleted because some literals in them are assigned 1. We need to prove that the remaining formula is good.
	\item R-Rule 3 is applied and we need to prove that the remaining formula is good.
\end{enumerate}

We will use the following two lemmas to help us solve these two cases.

\lem{gengood}{
	Let $\mathcal{F}$ be a formula containing a $(3^-,0^+)$ or $(0^+,2^-)$-literal $y$. Assume the total degree of $y$ is $a>0$.
	If we delete from $\mathcal{F}$ at most $a-1$ clauses and some literals other than $y$ and $\bar{y}$, where at least one deleted clause contains $y$, then the
	resulting formula is good.
}
\pf{
	Since the total degree of $y$ is $a$, at least one clause containing $y$ or $\bar{y}$ will not be deleted. Then $y$ or $\bar{y}$ will be a $(2^-,0^+)$-literal in the remaining formula.
	Thus the formula is good.
}\medskip

\crl{goodcond}{
	Let $\mathcal{F}$ be a reduced formula containing only $(3^-,3^-)$, $(2,4^+)$ and $(4^+,2)$-literals. For any literal $x$ in it with degree at most $4$, the formula $\mathcal{F}_x$ is good.
}
\pf{
	By \refl{lem-d4}, we know that the total degree of any literal in $\mathcal{F}$ is at least 5 and $\mathcal{F}$ does not contain any $1$-clauses.
	Note that $\mathcal{F}_x$ is obtained from $\mathcal{F}$ by deleting all clauses containing $x$ and deleting the literal $\bar {x}$.
	Any literal different from $x$ in a clause containing $x$ will be the literal $y$ in \refl{gengood}. By \refl{gengood}, we know the corollary holds.
}

\lem{resolutiontobegood}{
	Let $\mathcal{F}$ be a formula containing a $(1,1^+)$-literal $x$ and at least two different $(2^-,0^+)$-literals other than $x$ and $\bar{x}$.
	It holds that
	either $m_{\mathcal{F}_{\setminus x}} \leq m_{\mathcal{F}}-1$ and $\mathcal{F}_{\setminus x}$ is a good formula or
	$m_{\mathcal{F}_{\setminus x}} \leq m_{\mathcal{F}}-2$.
	% i.e., resolving on $x$ decreases the number of clauses by at least $2$.
}
\pf{
	Let the unique clause containing $x$ be $C$ and the clauses containing $\bar{x}$ be $D_1,D_2,\dots D_l$. Let $y$ and $z$ be two different $(2^-,0^+)$-literals other than $x$ and $\bar{x}$, where $y$ and $z$ can be each other's negation.
	
	It is easy to see that resolving on $x$ will decrease the number of clauses by at least $1$. We assume that the number of clauses decreases by exactly 1 after resolving on $x$ and show for this case the formula $\mathcal{F}_{\setminus x}$ must be good.
	For this case, the $l+1$ clauses  $C,D_1,D_2,\dots D_l$ are deleted and all the $l$ clauses $D_i\cup C\setminus \{x, \bar{x}\}$ $(i=1,2,\dots, l)$ are added in $\mathcal{F}_{\setminus x}$.
	
	Case 1. $x$ is a $(1,1)$-literal: after resolving on $x$, the degree of any literal does not increase and no literal other than $x$ and $\bar{x}$ disappears.
	So $y$ and $z$ are still $(2^-,0^+)$-literals, witnessing the goodness of $\mathcal{F}_{\setminus x}$.
	
	Case 2. $x$ is a $(1,2^+)$-literal: We further distinguish two cases: $|C|\geq 3$ and $|C|\leq 2$.
	If $|C|\geq 3$, then any pair of literals in $C\setminus\{x\}$ will be a coincident pair in $\mathcal{F}_{\setminus x}$. Thus, $\mathcal{F}_{\setminus x}$ is good.
	If $|C|\leq 2$, then at most one literal the degree of who will increase after resolving on $x$, since only the degree of literals in $C\setminus\{x\}$ will increase.
	So one of $y$ and $z$ will be remained as a $(2^-,0^+)$-literal in $\mathcal{F}_{\setminus x}$. Thus, $\mathcal{F}_{\setminus x}$ is good.
}

\section{Detailed branching operations}
In this section, we show the detailed branching operations in Algorithm~\ref{alg:algorithm}.
Recall that we only branch on reduced formulas.
The detailed branching steps for bad and good formulas are given in Sec.~\ref{sec_bad} and~\ref{sec_good}, respectively.
For a bad formula, if there exist $(3,4)$ or $(4,3)$-literals, then deal with them. Else we deal with $(3,3)$-literals. For a good formula, we first deal with $(3,5^+)$ or $(4^+,4^+)$-literals; second deal with
$(3,4)$-literals (and also $(4,3)$-literals); third deal with $(2,3^+)$-literals (and also $(3^+,2)$-literals); last there are only $(3,3)$-literals and we deal with them.

The main results of these steps are summarized in the following two tables, where the number with `$^*$' in the `Vectors' column means the corresponding branch will leave a good formula. From the two tables, we can see that direct analysis will get a bound of $O^*(1.2600^m)$ since the largest branching factor is 1.2600. This does not use amortization. Our deep analysis in the proof of \reft{thm-final} shows that we can improve the bound to $O^*(1.2226^m)$.

\begin{table}[h!]
	\caption{Branching for Bad Formulas}
	\centering
	\begin{tabular}{lccc}
		\hline
		Cases  & Literals & Vectors & Factors \\
		\hline
		Case 1   & $(3,4)$-literals  & (3,4)       & 1.2208   \\
		Case 2   & $(3,3)$-literals  & ($3^*,3^*$)   & 1.2600   \\
		\hline
	\end{tabular}
	\label{tab:badbranch}
\end{table}

\begin{table}[h!]
	\caption{Branching for Good Formulas}
	\centering
	\begin{tabular}{lccc}
		\hline
		Cases  & Literals & Vectors & Factors \\
		\hline
		Case 1   & ($3,5^+$)-literals      & (3,5)   & 1.1939   \\
		Case 1   & ($4^+,4^+$)-literals      & (4,4)   & 1.1893 \\
		Case 2   & ($3,4$)-literals      & (4,4)   & 1.1893 \\
		&                       & (3,5) or (5,3)   & 1.1939   \\
		&                       & ($3^*,4^*$) or ($4^*,3^*$)   & 1.2208   \\
		Case 3   & ($2,3^+$)-literals      & (4,4)   & 1.1893   \\
		&                       & (3,5) or (5,3)   & 1.1939   \\
		&                       & ($3^*,4^*$) or ($4^*,3^*$)   & 1.2208   \\
		Case 4   & ($3,3$)-literals      & (4,4)   & 1.1893   \\
		&                       & (3,5) or (5,3)   & 1.1939   \\
		&                       & ($3^*,4^*$) or ($4^*,3^*$)   & 1.2208   \\
		\hline
	\end{tabular}
	\label{tab:goodbranch}
\end{table}

\subsection{$\mathcal{F}$ is a bad formula} \label{sec_bad}
\paragraph*{Case 1.} $\mathcal{F}$ contains a $(3,4)$-literal $x$: We branch on $x$ into two branchings $\mathcal{F}_x$ and $\mathcal{F}_{\bar{x}}$.
The branching vector is $(3,4)$.

\paragraph*{Case 2.} $\mathcal{F}$ only contains $(3,3)$-literals: We branch on an arbitrary literal $x$ into two branchings $\mathcal{F}_x$ and $\mathcal{F}_{\bar{x}}$.
The branching vector is $(3,3)$. However, the two sub-instances in the two branchings are good formulas by \refc{goodcond}.

\subsection{$\mathcal{F}$ is a good formula} \label{sec_good}
\paragraph*{Case 1.} $\mathcal{F}$ contains a $(3,5^+)$ or $(4^+,4^+)$-literal $x$:
Branch on $x$ into two branchings $\mathcal{F}_x$ and $\mathcal{F}_{\bar{x}}$. The branching vector will be at least
$(3, 5)$ or $(4,4)$.

\paragraph*{Case 2.} $\mathcal{F}$ contains a $(3,4)$-literal (but no $(3,5^+)$ or $(4^+,4^+)$-literal):
We further distinguish several cases to analyze the branching vector.

\textbf{Case 2.1.} $\mathcal{F}$ also contains a $(2,3^+)$-literal $y$: We first branch on an arbitrary $(3,4)$-literal $x$ into two branchings $\mathcal{F}_x$ and $\mathcal{F}_{\bar{x}}$.
If there is a clause containing both $x$ and $y$, then in the branching $\mathcal{F}_x$, the degree of $y$ is at most 1. Thus
$y$ will become a $(1,1^+)$-literal or $(0,1^+)$-literal in $\mathcal{F}_x$ and we will further apply R-Rule 1 or 3 on $y$ to decrease
the number of clauses by at least 1. We can get a branching vector at least $(4,4)$.

If there is a clause containing both $\bar{x}$ and $y$, then in the branching $\mathcal{F}_{\bar{x}}$, the degree of $y$ is at most 1. We apply  R-Rule 1 or 3 on $y$ to further decrease the number of clauses by at least 1.
We can get a branching vector at least $(3,5)$.

The remaining case is that the clauses containing $x$ or $\bar{x}$ does not contain $y$.
For this case, we can only get a branching vector $(3,4)$. However, in each branching of $\mathcal{F}_x$ and $\mathcal{F}_{\bar{x}}$,
the new instance is a good formula, because there is at least one $(2,0^+)$-literal $y$ in them.

\textbf{Case 2.2.} $\mathcal{F}$ contains only $(3,4)$-literals, $(4,3)$-literals and $(3,3)$-literals:
Let $Y$ be the set of $(4,3)$-literals $x'$ such that there is a clause containing both $x'$ and a $(3,3^+)$-literal.
%$=\{x|x \text{ is a } (4,3)-literal \text{ and there is a clause containing both } x \text{ and a } (3,3^+)-literal\}$
%We branch on any literal $x'\in Y$ into two branchings $\mathcal{F}_{x'}$ and $\mathcal{F}_{\bar{x'}}$, which gives a branching vector $(4,3)$.

\textbf{Case 2.2.1.} $Y\neq \emptyset$: There is a literal $x\in Y$ and a clause containing $\bar{x}$ which does not contain any $(4,3)$-literals, otherwise R-Rule 5 could be applied and $\mathcal{F}$ would not be a reduced instance. Thus the clause containing $\bar{x}$ will contain some $(3,3^+)$-literals.
We branch on $x$ with a branching vector $(4,3)$.
By \refl{gengood}, we know that both branchings $\mathcal{F}_{x}$ and $\mathcal{F}_{\bar{x}}$ are good formulas.

\textbf{Case 2.2.2.} $Y= \emptyset$: For this case, $(4,3)$-literals appear in clauses containing only $(4,3)$-literals.
Now Conditions (1) and (4) in the definition of bad formulas hold. Since $\mathcal{F}$ is a good formula now, we know either Condition (2) or Condition (3) will not hold.
Thus there is either a $2$-clause or a coincident pair.

First, we assume that $\mathcal{F}$ contains a coincident pair $\{x, y\}$. If $x$ is a $(3,4)$-literal, then $y$ must be a $(3,3^+)$-literal. For this case, we branch on $x$ into two branchings $\mathcal{F}_{x}$ and $\mathcal{F}_{\bar{x}}$.
In the branching $\mathcal{F}_{x}$, literal $y$ becomes a $(1,1^+)$-literal or a $(0,1^+)$-literal and we can reduce the number of clauses by 1 by applying
R-Rule 3 or R-Rule 1 on $y$. We get a branching vector $(4,4)$.
If both of $x$ and $y$ are $(3,3)$-literals, we branch on an arbitrary $(3,4)$-literal with a branching vector $(3,4)$. Furthermore,
in each branching, the instance is a good formula because there is either a coincident pair $(x,y)$ or one of $x$ and $y$ becomes a literal of degree at most 2.
The remaining case is that both of $x$ and $y$ are $(4,3)$-literals. For this case, we branch on $x$ into two branchings $\mathcal{F}_{x}$ and $\mathcal{F}_{\bar{x}}$ with a branching vector $(4,3)$. The formula
$\mathcal{F}_{x}$ is good because literal $y$ becomes a $(2^-,1^+)$-literal. The formula $\mathcal{F}_{\bar{x}}$ is good by \refl{gengood}. Notice that for this case in $\mathcal{F}$ the clauses containing $\bar{x}$ cannot contain any $(4,3)$-literal and then each of them must contain another $(3,3^+)$-literal.

Second, we assume that $\mathcal{F}$ does not contain any coincident pair and there is a 2-clause $\{x,y\}$.
%If $x$ is a $(4,3)$-literal or $(3,4)$-literal contained in a 2-clause $C=\{x,y\}$,
We branch on $x$ into two branchings $\mathcal{F}_{x}$ and $\mathcal{F}_{\bar{x}}$. In the branching $\mathcal{F}_{\bar{x}}$, we get a 1-clause containing only $y$. Furthermore, $\mathcal{F}_{\bar{x}}$ has at least two clauses containing $y$ because $y$ and $\bar{x}$ do not form a coincident pair in $\mathcal{F}$. We apply R-Rule 1 on $y$ and can further decrease the number of clauses by at least 2. We get a branching vector at least $(3,5)$.

\paragraph*{Case 3.} $\mathcal{F}$ contains a $(2,3^+)$-literal (but no $(3,4^+)$ or $(4^+,3)$-literal):
Now $\mathcal{F}$ contains only $(2,3^+)$-literals, $(3^+,2)$-literals and $(3,3)$-literals.
We consider the following subcases.

\textbf{Case 3.1.} There is a $2$-clause $C=\{x,y\}$ containing a $(3^+,2^+)$-literal $x$:
%We will branch on $x$ into two branchings $\mathcal{F}_{x}$ and $\mathcal{F}_{\bar{x}}$.
We do a deeper analysis by considering different cases.

\textbf{Case 3.1.1.} Each clause containing $\bar{x}$ is a $2$-clause: We branch on $x$. In the branching of $\mathcal{F}_{x}$, we will get at least two 1-clauses. By applying R-Rule 1 on them, we can further reduce 2 clauses.
In the branching of $\mathcal{F}_{\bar{x}}$, we get at least one 1-clause. By applying R-Rule 1 on it, we can further reduce 1 clause. So we can get a branching vector $(5, 3)$ at least.

Next, we can assume that there is a literal $z\not\in \{x, \bar{x}, y, \bar{y}\}$ appearing in a clause containing $\bar{x}$.

\textbf{Case 3.1.2.} At least one of $x$ and $y$ is a $(3,3)$-literal: We assume that $x$ is a $(3,3)$-literal.
We branch on $x$.
In the branching of $\mathcal{F}_{\bar{x}}$, there is a 1-clause $\{y\}$ and we can reduce at least one clause by applying R-Rule 1 on it. Because $z$ exists, by \refl{gengood} we know that if only four clauses are removed in total, the remaining instance will be a good formula. In the branching of $\mathcal{F}_{x}$, three clauses are deleted and the remaining instance is also a good formula by \refc{goodcond}.
We can branch with a branching vector $(3, 4)$ with a good formula in each remaining branching or branch with a branching vector at least $(3,5)$.

\textbf{Case 3.1.3.} %Literal $x$ is a $(3^+,2)-literal$ and $y$ is a $(3^+,2^+)-literal$.
Both of $x$ and $y$ are $(3^+,2)$-literals: We further consider two subcases.

%If there is a clause contains both a literal from $\{x,y\}$ and a negation of a literal from $\{x,y\}$, we assume without loss of generality that there is a clause containing $x$ and $\bar{y}$. In the branching of

If each clause containing $\bar{x}$ also contains $y$, then we branch on $y$. In the branching of $\mathcal{F}_y$, literal $x$ will
become a $(2^+,0)$-literal. We can reduce two more clauses by applying R-Rule 1 on $x$.
In the branching of $\mathcal{F}_{\bar{y}}$, we will have a 1-clause $\{x\}$.  We can reduce at least one clause by applying R-Rule 1 on $\{x\}$. Then we can get a branching vector $(5,3)$ at least.

Otherwise, at most one clause containing $\bar{x}$ contains $y$. For this case, we branch on $x$.
In the branching of $\mathcal{F}_{\bar{x}}$, we will have a 1-clause $\{y\}$.  We can reduce at least two clauses by applying R-Rule 1 on $\{y\}$. As $z$ exists, by \refl{gengood} we know if just 4 clauses are removed in total, the remaining instance is a good formula.
For the branching of $\mathcal{F}_{x}$, three clauses are deleted and we can apply \refc{goodcond}. The remaining instance is also a good formula. So we get a branching vector $(3, 4)$ with a good formula in each remaining branching or a branching vector covered by $(3,5)$.

\textbf{Case 3.1.4.} Literal $x$ is a $(3^+,2)$-literal, $y$ is a $(2,3^+)$-literal,
and no clause contains both of $y$ and $\bar{x}$: We branch on $x$.
In the branching of $\mathcal{F}_x$,  literal $y$ will become a $(1^-,0^+)$-literal. We can reduce at least one clause by applying R-Rule 1 or R-Rule 3 on $y$. In the branching of $\mathcal{F}_{\bar{x}}$,  we will have a 1-clause $\{y\}$ and can reduce at least two clauses by applying R-Rule 1. Thus, we can get a branching vector of $(4,4)$.

\textbf{Case 3.1.5.} Literal $x$ is a $(3^+,2)$-literal, $y$ is a $(2,3^+)$-literal, and a clause contains both of $y$ and $\bar{x}$: We branch on $x$.

Assume that there is a $2$-clause other than $C$ containing $x$. In the branching of $\mathcal{F}_x$, we can further decrease the number of clauses by at least 1 by applying R-Rule 3 on $y$. In the branching of $\mathcal{F}_{\bar{x}}$, we will get at least two 1-clauses and can further decrease the number of clauses by at least 2
by applying R-Rule 1. We can get a branching vector covered by $(4,4)$.

Otherwise, the other two clauses containing $x$, denoted by $C_1$ and $C_2$, are both $3^+$-clauses.
We can simply assume that $C_i (i=1,2)$ does contains $C=(x,y)$ or both a literal and its negation, since for this case we can simply delete $C_i$ without branching.
Thus $C_1\cup C_2$ will contain at least two different literals $z_1$ and  $z_2$ that are also different from $x, \bar{x}, y$ and $\bar{y}$. If $x$ is a $(4^+,2)$-literal, in the branching of $\mathcal{F}_x$, we reduce at least four clauses directly and leave
a $(1,0^+)$-literal $y$. By applying R-Rule 1 or R-Rule 3 on $y$, we can further reduce at least one clause. So we can reduce
at least five clauses for this case. Next, we assume that $x$ is a $(3,2)$-literal. For this case,
in $\mathcal{F}_x$, literal $y$ will become a $(1,1^+)$-literal, and literals $z_1$ and $z_2$
will become two different $(2^-,0^+)$-literals (also different from $y$ and $\bar{y}$). By \refl{resolutiontobegood}, we know that after resolving $y$ in $\mathcal{F}_x$, we can reduce one clause with the resulting formula being good or
reduce at least two clauses directly. So in the branching of $\mathcal{F}_x$, we can either reduce four clauses leaving a good formula or reduce at least five clauses.
In the other branching of $\mathcal{F}_{\bar{x}}$, we get a 1-clause $\{y\}$, after applying R-Rule 1 on it we can further reduce
one clause. If only three clauses are reduced in this branching, then the remaining formula is good. The reason is as below.
In Case 3, $\mathcal{F}$ contains only $(2,3^+)$-literals, $(3^+,2)$-literals and $(3,3)$-literals.
There is a literal $z\not\in \{x, \bar{x}, y, \bar{y}\}$ appears in a clause containing $\bar{x}$ (after Case 3.1.1).
For this case, $z$ will be a $(2^-,0^+)$-literal or $(0^+,2)$-literal in the remaining formula and then the remaining formula is good.
We can branch with a branching vector $(4,3)$ leaving a good formula in each branching or a branching vector covered by $(5,3)$ or $(4,4)$.

\textbf{Case 3.2.} There is a $2$-clause $C=\{x,y\}$ containing two $(2,3^+)$-literals: We consider two subcases.

\textbf{Case 3.2.1.} There is no clause containing both of $y$ and $\bar{x}$: We branch on $x$.
In the branching of $\mathcal{F}_x$, literal $y$ will become a $(1^-,2^+)$-literal. We can reduce one more clause by applying R-Rule 3 on $y$. In the branching of  $\mathcal{F}_{\bar{x}}$, a 1-clause $\{y\}$ is created and there are two clauses containing $y$.
We can reduce two more clauses by applying R-Rule 1 on $y$. We get a branching vector of $(3,5)$.

\textbf{Case 3.2.2.} There is a clause $D$ containing both of $y$ and $\bar{x}$: If $D$ is also a 2-clause, then there are two 2-clauses $\{x,y\}$ and $\{\bar{x}, y\}$. We simply assign $y=1$ without branching. Next, we assume that $D$ is a $3^+$-clause.

If $D$ is a 3-clause, we branch on $y$. In the branching of $\mathcal{F}_y$, literal $x$ will become a $(1^-,2^+)$-literal. We can reduce one more clause by applying R-Rule 3 on $x$. In the branching of $\mathcal{F}_{\bar{y}}$, we will get two -clauses $\{x\}$ and $\{z\}$, where $z$ is the third literal in $D$. By applying R-Rule 1 on $\{x\}$ and $\{z\}$, we can reduce two more clauses.
We get a branching vector of $(3,5)$.

Else $D$ is a $4^+$-clause, and we branch on $x$.
In the branching of  $\mathcal{F}_x$, literal $y$ will become a $(1,2^+)$-literal. After applying R-Rule 3 on $y$, we reduce one more clause leaving a good formula, because $D$ contains at least two literals other than $y$ and $\bar{x}$ and then there is a coincident pair after applying R-Rule 3 on $y$.
In the branching of $\mathcal{F}_{\bar{x}}$, we will get a 1-clause $\{y\}$. We can reduce one more clause by applying R-Rule 1 on it. Same as before, if just 4 clauses are removed, the remaining instance is good. Thus, we can either get a branching vector $(3, 4)$ with a good formula in each remaining branching or a branching vector covered by $(3,5)$.

Next, we assume that there is no 2-clause.

\textbf{{Case 3.3.}} There is a clause in $\mathcal{F}$ containing both a $(3,3)$-literal $x$ and a $(2,3^+)$-literal $y$:
Let $C_1$, $C_2$ and $C_3$ be the three clauses containing $x$, where we assume that $C_1$ also contains $y$.
Let $C_4$ be the other clause containing $y$.
We first branch on $x$ with a branching vector $(3,3)$. We may decrease the number of clauses more
by applying reduction rules for different cases.

\textbf{Case 3.3.1.} $C_4=C_2$ or $C_4=C_3$: This means $\{x, y\}$ is a coincident pair. In the branching $\mathcal{F}_x$, the literal $y$ becomes a $(0, 2^+)$-literal. We can further remove at least two clauses by applying R-Rule 1 on $y$. We get a branching vector $(5,3)$. Next, we assume that $C_4\neq C_2$ or $C_3$.

\textbf{Case 3.3.2.} $C_4\neq C_2$ and  $C_4\neq C_3$:
Notice that $C_2$ and $C_3$ are $3^+$-clauses and each of them will contain a literal different from $\{x, \bar{x}, y ,\bar{y}\}$.
In $\mathcal{F}_x$,  there is a $(1,1^+)$-literal $y$ and two different $(2^-,0^+)$-literals different from $\{x, \bar{x}, y ,\bar{y}\}$. So it satisfies the condition in \refl{resolutiontobegood}.
After resolving $y$ in $\mathcal{F}_x$, we can further either reduce one clause leaving a good formula or reduce at least two clauses.
In the branching of $\mathcal{F}_{\bar{x}}$, we reduce three clauses directly and the remaining formula is good according to \refc{goodcond}.
So the branching vector is either $(4,3)$ with a good formula in each branching or a vector covered by $(5,3)$.

\lem{caseto}{
	For a reduced instance $\mathcal{F}$ without $(3^+,4^+)$-literals, if there is no 2-clause and no clause contains both a $(2,3^+)$-literal and a $(3,3)$-literal,
	then either there is no $(2,3^+)$-literal or there is a clause containing at least three $(2,3^+)$-literals.
}
\pf{
	Since $\mathcal{F}$ is a reduced instance, we know that the degree of any literal is at least 2 and there is no $(2,2)$-literal.
	Note that there is also no $(3^+,4^+)$-literal. Thus, the formula contains only $(2,3^+)$-literals, $(3^+,2)$-literals and $(3,3)$-literals. We assume that there is at least one $(2,3^+)$-literal otherwise the lemma trivially holds.
	It is impossible that each clause containing a $(2,3)$-literal also contains a $(3^+,2)$-literal because this case would be reduced
	by R-Rule 4. So there is a clause containing only $(2,3^+)$-literals. Since there is no 2-clause. We know that the clause contains
	at least three $(2,3^+)$-literals.
}

By \refl{caseto}, we know that the remaining case is as follows.

\textbf{Case 3.4.} There is a $3^+$-clause $C$ containing at least three $(2,3^+)$-literals $\{x_1,x_2,x_3\}$:
Let $C_i$ be the other clause containing $x_i$ ($i=1,2,3$), where it is possible two of $C_1$, $C_2$ and $C_3$ are the same.

\textbf{Case 3.4.1.} Two literals in $\{x_1,x_2,x_3\}$, say $x_1$ and $x_2$, form a coincident pair:
We branch on $x_1$ with a branching vector $(2,3)$ first. In the branching of $\mathcal{F}_{x_1}$, literal $x_2$ will become a
$(0,3^+)$-literal and we reduce three clauses by applying R-Rule 1 on $x_2$. So we can get a branching vector of $(5,3)$.

\textbf{Case 3.4.2.} At least one of $C_1,C_2$ and $C_3$ contains a negation of $x_1, x_2$ or $x_3$: Without loss of generality we assume that $C_2$ contains a negation of $x_1$.
We first branch on $x_1$ with a branching vector $(2,3)$. In the branching of $\mathcal{F}_{x_1}$, each of $x_2$ and $x_3$
will become a $(1,1^+)$-literal. We can further reduce the number of clauses by at least 2 by applying R-Rule 3 on $x_2$ and $x_3$ one by one. In the branching of $\mathcal{F}_{\bar{x_1}}$, after deleting the three clauses containing $\bar{x_1}$ (including $C_2$),
the degree of $x_2$ is at most 1. We can reduce one more clause by applying reduction rules on $x_2$. Thus, we can branch with a
branching vector $(4,4)$.

\textbf{Case 3.4.3.} None of Case 3.4.1 and Case 3.4.2 happens:
We first branch on $x_1$ with a branching vector $(2,3)$. In the branching of $\mathcal{F}_{x_1}$, each of $x_2$ and $x_3$
will become a $(1,3^+)$-literal. We can reduce two more clauses by applying R-Rule 3 on $x_2$ and $x_3$ one by one. Furthermore,
the remaining instance is a good formula, because applying R-Rule 3 will create coincident pairs in this case.
In the branching $\mathcal{F}_{\bar{x_2}}$, the formula is a good formula by \refc{goodcond}.
We get a branching vector $(3, 4)$ with a good formula in each branching.

\paragraph*{Case 4.} $\mathcal{F}$ contains only $(3,3)$-literals: Since $\mathcal{F}$ is a good formula, we know that there is either a coincident pair or a $2$-clause.

\textbf{Case 4.1.} $\mathcal{F}$ contains a coincident pair $\{x, y\}$: We branch on $x$ into two branchings $\mathcal{F}_{x}$ and $\mathcal{F}_{\bar{x}}$, and distinguish two subcases to analyze the branching operation.

\textbf{Case 4.1.1.} Three clauses contain $x$ and $y$ simultaneously:  In the branching of $\mathcal{F}_x$, the literal $y$ will become a $(0,3)$-literal and we can further decrease
the number of clauses by at least 3 by applying R-Rule 1. So we can get a branching vector  $(3,6)$ at least.

%\textbf{Case 4.1.2.} Only two clauses contain $x$ and $y$ simultaneously, and in $\mathcal{F}_x$ the clause containing $y$ is a $3^{+}$-clause: In the branching $\mathcal{F}_x$, the literal $y$ will become a $(1,2^+)$-literal and we can further decrease
%the number of clauses by 1 and create a coincident pair by applying R-Rule 3. Such the remaining formula is good.
%In the branching $\mathcal{F}_{\bar{x}}$,  3 clauses are removed and at least one $(2^-,3^-)$-literal is created. So
%$\mathcal{F}_{\bar{x}}$ is also a good formula. We can always branch with a branching vector $(4,3)$ such that both sub instances
%are good formulas.

\textbf{Case 4.1.2.} Only two clauses contain $x$ and $y$ simultaneously: we assume without loss of generality that no pair of literals appear in more than two clauses simultaneously now.

Assume that one of the clauses containing $x$ is a $2$-clause $\{x,w\}$, where $w$ can be $y$.
In the branching of $\mathcal{F}_x$, we can apply R-Rule 3 on $y$ to further reduce 1 clause.
In the branching of $\mathcal{F}_{\bar{x}}$, we can apply R-Rule 1 on $w$ to further reduce 1 clause.
The branching vector will be covered by $(4,4)$.

Next, we assume that any of the three clauses containing $x$ also contains a literal other than $y$ and $\bar{y}$.
At least two of the three literals are different because no pair of literals appear in three clauses as assumed.
Let $z_1$ and $z_2$ be the two different literals. In $\mathcal{F}_x$, literal $y$ will become a $(1,1^+)$-literal and
$z_1$ and $z_2$ will become $(2^-,0^+)$-literals. The condition in \refl{resolutiontobegood} holds.
After resolving $y$ in $\mathcal{F}_x$, we can further either reduce 1 clause leaving a good formula or reduce at least 2 clauses.
In the branching of $\mathcal{F}_{\bar{x}}$, we reduce three clauses directly and the leaving formula is good according to \refc{goodcond}. The branching vector is either $(4,3)$ with a good formula in each branching or a vector covered by $(5,3)$.

%We assume that clause $C_1$ and $C_2$ contain both of $x$ and $y$, $C_3$ contains $x$ but not $y$, and $C_4$ contains $y$ but not $x$.
%We can also assume that the length of $C_3$ is not less than the length of $C_4$. Since $C_1$ and $C_2$ are different, at least one of
%them, say $C_1$ contains at least one literal different from $x$ and $y$. In the branching $\mathcal{F}_x$, literal $y$ becomes a $(1,2^+)$-literal and we further apply R-Rule 3 on $y$. Let $M$ be the set of literals in $C_1, C_2$ and $C_3$ other than $x$ and $y$. Then the degree of any literal in $M$ will decrease by at least 1 after assigning $x=1$. Let $M'$ be the set of literals in $C_4$ other than $y$, the degree of any literal in $M'$ will increase after applying R-Rule 3 on $y$. However, we have that $|M|> |M'|$ since the length of $C_3$ is not less then the length of $C_4$ and $C_1$ contains at least one literal different from $x$ and $y$. Thus, in this branching, we can create at least one $(2^-,3^-)$-literal and then it is a good formula.
%In the other branching $\mathcal{F}_{\bar{x}}$, three clauses are removed and by \refc{} it is still a good formula.

\textbf{Case 4.2.} $\mathcal{F}$ does not contain a coincident pair but contains a 2-clause $\{x, y\}$:
We branch on $x$ with a branching vector $(3,3)$. In the branching $\mathcal{F}_{\bar{x}}$, we will get a $1$-clause that only contains $y$. Furthermore, since $\mathcal{F}$ does not contain a coincident pair, we know that there are at least two clauses containing $y$ in $\mathcal{F}_{\bar{x}}$. We can apply R-Rule 1 on $y$ in $\mathcal{F}_{\bar{x}}$ to further reduce 2 clauses. Thus, we can get a branching vector covered by $(3,5)$.

\section{Conclusion}
SAT is one of the most widely studied NP-complete problems. There is a large number of references in the history, whether from the perspective of experimental algorithms or theoretical algorithms.
Many fast solvers have been developed and they can solve medium-large sized instances within a reasonable running time bound.
However, the theoretical research is relatively backward. It took us decades to improve the running time bound to $O^*(1.2226^m)$.
%For SAT measured by the number $n$ of clauses, most previous algorithms, 
%such as the algorithms in \cite{BetterGeneral,PPZ,19kSAT}, use randomization techniques.
%However, the best bound is still the trivial bound of $O^*(2^n)$ and the nonexistence of better bounds 
%is known as the strong exponential time hypothesis.
%The amortized technique used in this paper is simple but can be used to analyze 
%many branch-and-search algorithms, and may also provide a way for studying
%the strong exponential time hypothesis.
According to the theoretical results, the size of the problems we can solve is much smaller than that of the problems solved by fast practical solvers.
The gap between theoretical and experimental results is large. It is interesting to further explore the problem nature and reduce the gap, especially to accelerate the research of theoretical algorithms and explain the fast experimental algorithms.


\begin{thebibliography}{10}

	\bibitem{Chen}
Jianer Chen and Yang Liu.
\newblock An improved {SAT} algorithm in terms of formula length.
\newblock In {\em Algorithms and Data Structures, 11th International Symposium,
	{WADS} 2009, Banff, Canada, August 21-23, 2009. Proceedings}, pages 144--155,
2009.
\newblock URL: \url{https://doi.org/10.1007/978-3-642-03367-4\_13}, \href
{http://dx.doi.org/10.1007/978-3-642-03367-4\_13}
{\path{doi:10.1007/978-3-642-03367-4\_13}}.

\bibitem{COOK1971}
Stephen~A. Cook.
\newblock The complexity of theorem-proving procedures.
\newblock In {\em Proceedings of the 3rd Annual {ACM} Symposium on Theory of
	Computing, May 3-5, 1971, Shaker Heights, Ohio, {USA}}, pages 151--158, 1971.
\newblock URL: \url{https://doi.org/10.1145/800157.805047}, \href
{http://dx.doi.org/10.1145/800157.805047} {\path{doi:10.1145/800157.805047}}.

\bibitem{DantsinFirst}
Evgeny Dantsin.
\newblock Two systems for proving tautologies, based on the split method.
\newblock {\em Journal of Mathematical Sciences}, 22:1293--1305, 06 1983.
\newblock \href {http://dx.doi.org/10.1007/BF01084392}
{\path{doi:10.1007/BF01084392}}.

\bibitem{Schoning}
Evgeny Dantsin, Andreas Goerdt, Edward~A. Hirsch, Ravi Kannan, Jon~M.
Kleinberg, Christos~H. Papadimitriou, Prabhakar Raghavan, and Uwe
Sch{\"{o}}ning.
\newblock A deterministic (2-2/(k+1))\({}^{\mbox{n}}\) algorithm for k-sat
based on local search.
\newblock {\em Theor. Comput. Sci.}, 289(1):69--83, 2002.
\newblock URL: \url{https://doi.org/10.1016/S0304-3975(01)00174-8}, \href
{http://dx.doi.org/10.1016/S0304-3975(01)00174-8}
{\path{doi:10.1016/S0304-3975(01)00174-8}}.

\bibitem{FirstGeneral}
Evgeny Dantsin, Edward~A. Hirsch, and Alexander Wolpert.
\newblock Algorithms for {SAT} based on search in hamming balls.
\newblock In {\em {STACS} 2004, 21st Annual Symposium on Theoretical Aspects of
	Computer Science, Montpellier, France, March 25-27, 2004, Proceedings}, pages
141--151, 2004.
\newblock URL: \url{https://doi.org/10.1007/978-3-540-24749-4\_13}, \href
{http://dx.doi.org/10.1007/978-3-540-24749-4\_13}
{\path{doi:10.1007/978-3-540-24749-4\_13}}.

\bibitem{BetterGeneral}
Evgeny Dantsin and Alexander Wolpert.
\newblock Derandomization of schuler's algorithm for {SAT}.
\newblock {\em Electronic Colloquium on Computational Complexity {(ECCC)}},
(017), 2004.
\newblock URL:
\url{http://eccc.hpi-web.de/eccc-reports/2004/TR04-017/index.html}.

\bibitem{Davis}
Martin Davis and Hilary Putnam.
\newblock A computing procedure for quantification theory.
\newblock {\em J. {ACM}}, 7(3):201--215, 1960.
\newblock URL: \url{http://doi.acm.org/10.1145/321033.321034}, \href
{http://dx.doi.org/10.1145/321033.321034} {\path{doi:10.1145/321033.321034}}.

\bibitem{exactalgorithm}
Fedor~V. Fomin and Dieter Kratsch.
\newblock {\em Exact Exponential Algorithms}.
\newblock Texts in Theoretical Computer Science. An {EATCS} Series. Springer,
2010.
\newblock URL: \url{https://doi.org/10.1007/978-3-642-16533-7}, \href
{http://dx.doi.org/10.1007/978-3-642-16533-7}
{\path{doi:10.1007/978-3-642-16533-7}}.

\bibitem{AGuide1979}
M.~R. Garey and David~S. Johnson.
\newblock {\em Computers and Intractability: {A} Guide to the Theory of
	NP-Completeness}.
\newblock W. H. Freeman, 1979.

\bibitem{Van}
Allen~Van Gelder.
\newblock A satisfiability tester for non-clausal propositional calculus.
\newblock {\em Inf. Comput.}, 79(1):1--21, 1988.
\newblock URL: \url{https://doi.org/10.1016/0890-5401(88)90014-4}, \href
{http://dx.doi.org/10.1016/0890-5401(88)90014-4}
{\path{doi:10.1016/0890-5401(88)90014-4}}.

\bibitem{19kSAT}
Thomas~Dueholm Hansen, Haim Kaplan, Or~Zamir, and Uri Zwick.
\newblock Faster \emph{k}-sat algorithms using biased-ppsz.
\newblock In {\em Proceedings of the 51st Annual {ACM} {SIGACT} Symposium on
	Theory of Computing, {STOC} 2019, Phoenix, AZ, USA, June 23-26, 2019}, pages
578--589, 2019.
\newblock URL: \url{https://doi.org/10.1145/3313276.3316359}, \href
{http://dx.doi.org/10.1145/3313276.3316359}
{\path{doi:10.1145/3313276.3316359}}.

\bibitem{Hirsch}
Edward~A. Hirsch.
\newblock Two new upper bounds for {SAT}.
\newblock In {\em Proceedings of the Ninth Annual {ACM-SIAM} Symposium on
	Discrete Algorithms, 25-27 January 1998, San Francisco, California, {USA}},
pages 521--530, 1998.
\newblock URL: \url{http://dl.acm.org/citation.cfm?id=314613.314838}.

\bibitem{Hirsch2}
Edward~A. Hirsch.
\newblock New worst-case upper bounds for {SAT}.
\newblock {\em J. Autom. Reasoning}, 24(4):397--420, 2000.
\newblock URL: \url{https://doi.org/10.1023/A:1006340920104}, \href
{http://dx.doi.org/10.1023/A:1006340920104}
{\path{doi:10.1023/A:1006340920104}}.

\bibitem{SETH}
Russell Impagliazzo and Ramamohan Paturi.
\newblock On the complexity of k-sat.
\newblock {\em J. Comput. Syst. Sci.}, 62(2):367--375, 2001.
\newblock URL: \url{https://doi.org/10.1006/jcss.2000.1727}, \href
{http://dx.doi.org/10.1006/jcss.2000.1727}
{\path{doi:10.1006/jcss.2000.1727}}.

\bibitem{Kullmann}
O.~Kullmann.
\newblock Deciding propositional tautologies: Algorithms and their complexity.
\newblock 09 1997.

\bibitem{Monien1980}
Burkhard Monien and Ewald Speckenmeyer.
\newblock Upper bounds for covering problems.
\newblock {\em Methods of Operations Research.}, 43, 01 1980.

\bibitem{MonienFirst}
Burkhard Monien and Ewald Speckenmeyer.
\newblock Solving satisfiability in less than 2\({}^{\mbox{n}}\) steps.
\newblock {\em Discrete Applied Mathematics}, 10(3):287--295, 1985.
\newblock URL: \url{https://doi.org/10.1016/0166-218X(85)90050-2}, \href
{http://dx.doi.org/10.1016/0166-218X(85)90050-2}
{\path{doi:10.1016/0166-218X(85)90050-2}}.

\bibitem{Niedermeier}
Rolf Niedermeier and Peter Rossmanith.
\newblock An efficient fixed-parameter algorithm for 3-hitting set.
\newblock {\em Journal of Discrete Algorithms}, 1(1):89 -- 102, 2003.
\newblock Combinatorial Algorithms.
\newblock URL:
\url{http://www.sciencedirect.com/science/article/pii/S1570866703000091},
\href {http://dx.doi.org/https://doi.org/10.1016/S1570-8667(03)00009-1}
{\path{doi:https://doi.org/10.1016/S1570-8667(03)00009-1}}.

\bibitem{PPZ}
Ramamohan Paturi, Pavel Pudl{\'{a}}k, and Francis Zane.
\newblock Satisfiability coding lemma.
\newblock In {\em 38th Annual Symposium on Foundations of Computer Science,
	{FOCS} '97, Miami Beach, Florida, USA, October 19-22, 1997}, pages 566--574,
1997.
\newblock URL: \url{https://doi.org/10.1109/SFCS.1997.646146}, \href
{http://dx.doi.org/10.1109/SFCS.1997.646146}
{\path{doi:10.1109/SFCS.1997.646146}}.

\bibitem{JARobinson}
John~Alan Robinson.
\newblock A machine-oriented logic based on the resolution principle.
\newblock {\em J. {ACM}}, 12(1):23--41, 1965.
\newblock URL: \url{http://doi.acm.org/10.1145/321250.321253}, \href
{http://dx.doi.org/10.1145/321250.321253} {\path{doi:10.1145/321250.321253}}.

\bibitem{Schuler}
Rainer Schuler.
\newblock An algorithm for the satisfiability problem of formulas in
conjunctive normal form.
\newblock {\em J. Algorithms}, 54(1):40--44, 2005.
\newblock URL: \url{https://doi.org/10.1016/j.jalgor.2004.04.012}, \href
{http://dx.doi.org/10.1016/j.jalgor.2004.04.012}
{\path{doi:10.1016/j.jalgor.2004.04.012}}.

%\bibitem{wahlstrom2}
%Magnus Wahlström.
%\newblock Exact algorithms for finding minimum transversals in rank-3
%hypergraphs.
%\newblock {\em Journal of Algorithms}, 51(2):107 -- 121, 2004.
%\newblock URL:
%\url{http://www.sciencedirect.com/science/article/pii/S0196677404000136},
%\href {http://dx.doi.org/https://doi.org/10.1016/j.jalgor.2004.01.001}
%{\path{doi:https://doi.org/10.1016/j.jalgor.2004.01.001}}.

\bibitem{Wahlstrom}
Magnus Wahlstr{\"{o}}m.
\newblock Faster exact solving of {SAT} formulae with a low number of
occurrences per variable.
\newblock In {\em Theory and Applications of Satisfiability Testing, 8th
	International Conference, {SAT} 2005, St. Andrews, UK, June 19-23, 2005,
	Proceedings}, pages 309--323, 2005.
\newblock URL: \url{https://doi.org/10.1007/11499107\_23}, \href
{http://dx.doi.org/10.1007/11499107\_23} {\path{doi:10.1007/11499107\_23}}.

\bibitem{Yamamoto}
Masaki Yamamoto.
\newblock An improved o(1.234\({}^{\mbox{m}}\))-time deterministic algorithm
for {SAT}.
\newblock In {\em Algorithms and Computation, 16th International Symposium,
	{ISAAC} 2005, Sanya, Hainan, China, December 19-21, 2005, Proceedings}, pages
644--653, 2005.
\newblock URL: \url{https://doi.org/10.1007/11602613\_65}, \href
{http://dx.doi.org/10.1007/11602613\_65} {\path{doi:10.1007/11602613\_65}}.



\end{thebibliography}
\end{document}